\documentclass[a4paper]{spie} 
\usepackage[]{graphicx}

\title{The ASTRO-H Mission} 

\normalsize
\author{Tadayuki Takahashi\supit{a},
Kazuhisa Mitsuda\supit{a}, 
Richard Kelley\supit{b},
Felix~Aharonian\supit{c},\\
Fumie~Akimoto\supit{d},
Steve~Allen\supit{e},
Naohisa~Anabuki\supit{f},
Lorella~Angelini\supit{b},
Keith~Arnaud\supit{g},
Hisamitsu~Awaki\supit{h},
Aya~Bamba\supit{c},
Nobutaka~Bando\supit{a},
Mark~Bautz\supit{i},
Roger~Blandford\supit{e},
Kevin~Boyce\supit{b},
Greg~Brown\supit{j},
Maria~Chernyakova\supit{c},
Paolo~Coppi\supit{k},
Elisa~Costantini\supit{l},
Jean~Cottam\supit{b},
John~Crow\supit{b},
Jelle~de~Plaa\supit{l},
Cor~de~Vries\supit{l},
Jan-Willem~den Herder\supit{l},
Michael~DiPirro\supit{b},
Chris~Done\supit{m},
Tadayasu~Dotani\supit{a},
Ken~Ebisawa\supit{a},
Teruaki~Enoto\supit{e},
Yuichiro~Ezoe\supit{n},
Andrew~Fabian\supit{o},
Ryuichi~Fujimoto\supit{p},
Yasushi~Fukazawa\supit{q},
Stefan~Funk\supit{e},
Akihiro~Furuzawa\supit{d},
Massimiliano~Galeazzi\supit{r},
Poshak~Gandhi\supit{a},
Keith~Gendreau\supit{b},
Kirk~Gilmore\supit{e},
Yoshito~Haba\supit{d},
Kenji~Hamaguchi\supit{g},
Isamu~Hatsukade\supit{s},
Kiyoshi~Hayashida\supit{f},
Junko~Hiraga\supit{t},
Kazuyuki~Hirose\supit{a},
Ann~Hornschemeier\supit{b},
John~Hughes\supit{u},
Una~Hwang\supit{v},
Ryo~Iizuka\supit{w},
Kazunori~Ishibashi\supit{d}, 
Manabu~Ishida\supit{a},
Kosei~Ishimura\supit{a},
Yoshitaka~Ishisaki\supit{n},
Naoki~Isobe\supit{x},
Masayuki~Ito\supit{y},
Naoko~Iwata\supit{a},
Jelle~Kaastra\supit{l},
Timothy~Kallman\supit{b},
Tuneyoshi~Kamae\supit{e},
Hideaki~Katagiri\supit{q},
Jun~Kataoka\supit{z},
Satoru~Katsuda\supit{b},
Madoka~Kawaharada\supit{a},
Nobuyuki~Kawai\supit{aa},
Shigeo~Kawasaki\supit{a},
Dmitry~Khangaluyan\supit{a},
Caroline~Kilbourne\supit{b},
Kenzo~Kinugasa\supit{ab},
Shunji~Kitamoto\supit{ac},
Tetsu~Kitayama\supit{ad},
Takayoshi~Kohmura\supit{ae},
Motohide~Kokubun\supit{a},
Tatsuro~Kosaka\supit{af},
Taro~Kotani\supit{ag},
Katsuji~Koyama\supit{ah},  
Aya~Kubota\supit{ai},
Hideyo Kunieda\supit{d}, 
Philippe~Laurent\supit{aj},
Fran\c{c}ois~Lebrun\supit{aj},\\
Olivier~Limousin\supit{aj},
Michael~Loewenstein\supit{b},
Knox~Long\supit{ak},
Grzegorz~Madejski\supit{e},
Yoshitomo~Maeda\supit{a},
Kazuo Makishima\supit{t}, 
Maxim~Markevitch\supit{al},
Hironori~Matsumoto\supit{d},
Kyoko~Matsushita\supit{am},
Dan~McCammon\supit{an},
Jon~Miller\supit{ao},
Shin~Mineshige\supit{x},
Kenji~Minesugi\supit{a},
Takuya~Miyazawa\supit{d},
Tsunefumi~Mizuno\supit{q},
Koji~Mori\supit{s},
Hideyuki~Mori\supit{a},
Koji~Mukai\supit{b},
Hiroshi~Murakami\supit{ac},
Toshio~Murakami\supit{p},
Richard~Mushotzky\supit{g},
Yujin~Nakagawa\supit{ap},
Takao~Nakagawa\supit{a},
Hiroshi~Nakajima\supit{f},
Takeshi~Nakamori\supit{z},
Kazuhiro~Nakazawa\supit{t},
Yoshiharu~Namba\supit{aq},
Masaharu~Nomachi\supit{ar},
Steve~O'~Dell\supit{aw},
Hiroyuki~Ogawa\supit{a},
Mina~Ogawa\supit{a},
Keiji~Ogi\supit{as},
Takaya~Ohashi\supit{n},
Masanori~Ohno\supit{a},
Masayuki~Ohta\supit{a},
Takashi~Okajima\supit{v},
Naomi~Ota\supit{am},
Masanobu~Ozaki\supit{a},
Frits~Paerels\supit{at},
St$\acute{\rm e}$phane~Paltani\supit{au},
Arvind~Parmer\supit{av}, 
Robert~Petre\supit{b}, 
Martin~Pohl\supit{au},
Scott~Porter\supit{b},
Brian~Ramsey\supit{aw},
Christopher~Reynolds\supit{g},
Shin-ichiro~Sakai\supit{a},
Rita~Sambruna\supit{b},
Goro~Sato\supit{a},
Yoichi~Sato\supit{ax},
Peter~Serlemitsos\supit{b},
Maki~Shida\supit{a},
Takanobu~Shimada\supit{a},
Keisuke~Shinozaki\supit{ax},
Peter~Shirron\supit{b},
Randall~Smith\supit{al},
Gary~Sneiderman\supit{b},
Yang~Soong\supit{b},
Lukasz~Stawarz\supit{a},
Hiroyuki~Sugita\supit{ax},
Andrew~Szymkowiak\supit{k},
Hiroyasu~Tajima\supit{e},
Hiromitsu~Takahashi\supit{q},
Yoh~Takei\supit{a},
Toru~Tamagawa\supit{ap},
Takayuki~Tamura\supit{a},
Keisuke~Tamura\supit{a},
Takaaki~Tanaka\supit{e},
Yasuo~Tanaka\supit{a},\\
Yasuyuki~Tanaka\supit{a},
Makoto~Tashiro\supit{ay},
Yuzuru~Tawara\supit{d},
Yukikatsu~Terada\supit{ay},
Yuichi~Terashima\supit{h},
Francesco~Tombesi\supit{b},
Hiroshi~Tomida\supit{a},
Miyako~Tozuka\supit{am},
Yoko~Tsuboi\supit{w},
Masahiro~Tsujimoto\supit{a},
Hiroshi~Tsunemi\supit{f},
Takeshi~Tsuru\supit{ah}, \\
Hiroyuki~Uchida\supit{f},
Yasunobu~Uchiyama\supit{e},
Hideki~Uchiyama\supit{t},
Yoshihiro~Ueda\supit{x},
Shinichiro~Uno\supit{az},
Meg Urry\supit{ba},
Shin~Watanabe\supit{a},
Nicholas White\supit{b},
Takahiro~Yamada\supit{a},
Hiroya~Yamaguchi\supit{ap},
Kazutaka~Yamaoka\supit{ag},
Noriko~Yamasaki\supit{a},
Makoto~Yamauchi\supit{s},
Shigeo~Yamauchi\supit{bb},
Yoichi~Yatsu\supit{aa},
Daisuke~Yonetoku\supit{p},
Atsumasa~Yoshida\supit{ag}
 \skiplinehalf
\supit{a} Institute of Space and Astronautical Science (ISAS), JAXA, Kanagawa, 252-5210, Japan; 
\supit{b} NASA/Goddard Space Flight Center, Greenbert, MD 20771, USA;
\supit{c}Dublin Institute for Advanced Studies, Dublin 2, Ireland;
\supit{d}Department of Physics, Nagoya University, Nagoya,  338-8570, Japan;
\supit{e}Kavli Institute for Particle Astrophysics and Cosmology, Stanford University, CA 94305, USA;
\supit{f}Department of Earth and Space Science, Osaka University, Osaka,  560-0043, Japan;
\supit{g}Department of Physics, University of Maryland, MD 21250, USA;
\supit{h}Department of Physics, Ehime University, Ehime, 790-8577, Japan; 
\supit{i}Kavli Institute for Astrophysics and Space Research, Massachusetts Institute of Technology,  Cambridge, MA 02139, USA;
\supit{j}Lawrence Livermore National Laboratory, CA, 94550, USA;\\
\supit{k}Department of Physics, Yale University,  CT 06520-8120, USA;
\supit{l}SRON Netherlands Institute for Space Research, Utrecht, the Netherlands;
\supit{m}Department of Physics, University of Durham,  DH1 3LE, UK;
\supit{n}Department of Physics, Tokyo Metropolitan University, Tokyo, 192-0397, Japan;
\supit{o}Institute of Astronomy, Cambridge University, CB3 0HA, UK;
\supit{p}Faculty of Mathematics and Physics, Kanazawa University, Ishikawa,  920-1192, Japan;
\supit{q}Department of Physical Science, Hiroshima University, Hiroshima , 739-8526, Japan;
\supit{r}Physics Department, University of Miami, FL 33124, USA;
\supit{s}Department of Applied Physics, University of Miyazaki, Miyazaki, 889-2192, Japan;
\supit{t}Department of Physics, University of Tokyo, Tokyo,  113-0033, Japan;
\supit{u}Department of Physics and Astronomy, Rutgers University, NJ 08854-8019, USA;
\supit{v}Department of Physics and Astronomy, Johns Hopkins University, MD 21218, USA;
\supit{w}Department of Physics, Chuo University, Tokyo 112-8551, Japan;
\supit{x}Department of Astronomy, Kyoto University, Kyoto 606-8502, Japan;
\supit{y}Faculty of Human Development, Kobe University, Hyogo, 657-8501, Japan;
\supit{z}Research Institute for Science and Engineering, Waseda University, Tokyo 169-8555, Japan;
\supit{aa}Department of Physics, Tokyo Institute of Technology, Tokyo 152-8551, Japan;
\supit{ab}Gunma Astronomical Observatory, Gunma 377-0702, Japan;
\supit{ac}Department of Physics, Rikkyo University,Tokyo 171-8501, Japan;
\supit{ad}Department of Physics, Toho University, Chiba 274-8510, Japan;
\supit{ae}Department of Physics, Kougakuin University, Hachioji, Tokyo 192-0015, Japan;
\supit{af}School of Systems Engineering, Kochi University of Technology,  Kochi 782-8502, Japan;
\supit{ag}Department of Physics and Mathematics, Aoyama Gakuin University,  Kanagawa 229-8558, Japan;
\supit{ah}Department of Physics, Kyoto University, Kyoto,  606-8502, Japan;
\supit{ai}Department of Electronic Information Systems, Shibaura Institute of Technology, Saitama 337-8570, Japan;
\supit{aj}IRFU/Service d'Astrophysique, CEA Saclay, 91191 Gif-sur-Yvette, Cedex France;
\supit{ak}Space Science Telescope Institute, Baltimore, MD 21218 USA;
\supit{al}Harvard-Smithsonian Center for Astrophysics, MA 02138, USA;
\supit{am}Department of Physics, Tokyo University of Science, Tokyo, 162-8601, Japan;
\supit{an}Department of Physics, University of Wisconsin, WI 53706, USA;
\supit{ao}Department of Astronomy, University of Michigan, Ann Arbor, MI 48109;
\supit{ap}RIKEN, Saitama 351-0198, Japan;
\supit{aq}Department of Mechanical Engineering, Chubu University, Kasugai, Aichi 487-8501, Japan;
\supit{ar}Laboratory of Nuclear Studies, Osaka University, Osaka,  560-0043, Japan;
\supit{as}Graduate School of Science and Engineering,Ehime University, Ehime, 790-8577, Japan; 
\supit{at}Columbia Astrophysics Laboratory, Department of Astronomy, Columbia University,  NY 10027, USA
\supit{au}Universit$\acute{\rm e}$ de Gen\`eve, Switzerland;
\supit{av}ESTEC, Space Science Dept., Noordwijk, The Netherlands;
\supit{aw}NASA/Marshall Space Flight Center, AL 35812, USA;
\supit{ax}Aerospace Research and Development Directorate, JAXA, Ibaraki, 305-8505, Japan;
\supit{ay}Department of Physics, Saitama University, Saitama,  338-8570, Japan;
\supit{az}Faculty of Social and Information Sciences, Nihon Fukushi University, Aichi 475-0012, Japan;
\supit{ba}Yale Center for Astronomy and Astrophysics, Yale University, New Haven, CT 06520-8121, USA;
\supit{bb}Department of Physics, Faculty of Science, Nara Women's University, Nara, Nara 630-8506, Japan;}

 \authorinfo{ASTRO-H Web site: http://ASTRO-H.isas.jaxa.jp/index.html.en}

\begin{document} 
  \maketitle 

\begin{abstract}

The joint JAXA/NASA ASTRO-H  mission is the sixth in a series of highly successful 
X-ray missions initiated by the Institute of Space and Astronautical Science (ISAS). 
ASTRO-H will investigate the physics of the high-energy universe by performing 
high-resolution, high-throughput spectroscopy with moderate angular resolution.
ASTRO-H covers very wide 
energy range from 0.3~keV to 600~keV.  ASTRO-H allows a combination of wide band 
X-ray spectroscopy (5--80~keV) provided by multilayer coating, focusing hard 
X-ray mirrors and hard X-ray imaging detectors, and high energy-resolution soft 
X-ray spectroscopy (0.3--12~keV) provided by thin-foil X-ray optics and a 
micro-calorimeter array.  The mission will also carry an X-ray CCD camera 
as a focal plane detector for a soft X-ray telescope  (0.4--12~keV) and a  non-focusing 
soft gamma-ray detector  (40--600~keV) . The micro-calorimeter system is 
developed by an international collaboration led by ISAS/JAXA and NASA. 
The simultaneous broad bandpass, coupled with high spectral resolution 
of $\Delta E $ $\sim$7~eV provided by the micro-calorimeter will enable 
a wide variety of important science themes to be pursued.
\end{abstract}




\section{Introduction}

ASTRO-H, which was formerly called NeXT, is an international X-ray satellite 
that Japan plans to launch with the H-II A rocket in 2014\cite{Ref:Proposal,Ref:Proposal03,Ref:Kunieda2004,Ref:Takahashi,Ref:Takahashi2008}.  
NASA has selected the US participation in  ASTRO-H as a Mission of Opportunity.   
Under this program, the NASA/Goddard Space Flight Center collaborates with 
ISAS/JAXA on the implementation of an X-ray micro-calorimeter spectrometer
(SXS Proposal NASA/GSFC, 2007)\cite{Ref:ProposalNASA}. Other international members are from Stanford 
University, SRON, Geneva University and CEA/DSM/IRFU. In addition, in early 
2009, NASA, ESA and JAXA have selected science working group members to provide 
scientific guidance to the ASTRO-H project relative to the design/development 
and operation phases of the mission.

The history and evolution of the Universe can be described as a process in which structures 
of different scales are formed such as stars, galaxies, and clusters of galaxies, 
but also as matter and energy concentrate to an extreme degree in a form 
of black holes and neutron stars.  It is a mysery of Nature that the overwhelming 
diversity over tens or orders in spatial and density scales has been produced 
in the Universe following an expansion from a nearly uniform state. Clusters of galaxies
are the largest astronomical object in the Universe. Observing clusters of galaxies and revealing
their history is bound to lead to an understanding of how the structures are formed and 
evolved in the Universe. Equally important is to study how galaxies and supermassive black holes
form and develop,  and what role they play in forming clusters of galaxies.


About 80\% of the observable matter in the Universe is thought be in a form of hot gas that can 
be directly studied only in the X-ray band. X-ray observations are indispensable for 
unveiling the mysteries of the Universe. The ASTRO-H mission is equipped with a suite of 
instruments with the highest energy resolution ever achieved at $E >$ 3 keV and a wide energy range 
spanning four decades in energy from soft X-rays to gamma-rays. The mission aims to 
understand the dynamics of the evolution of the Universe and the concentration of the 
energy including how the most energetic particles, that are still far from thermal equilibrium, are produced.  

\begin{figure}
\centerline{\includegraphics[width=8cm,angle=0]{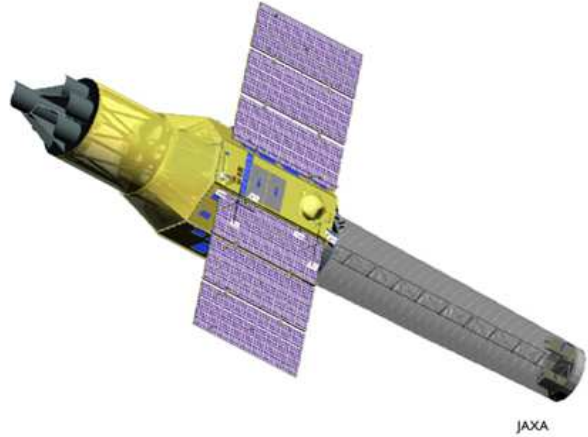}}
\caption{Artist's drawing of the ASTRO-H satellite. The focal  length of the Hard X-ray Telescope (HXT) is 12m, whereas the Soft X-ray Spectrometer (SXS) and Soft X-ray Imager (SXI)
 will have a focal length of 5.6~meters.}
\label{Fig:ASTRO-H}
\end{figure}

ASTRO-H has just completed the preliminary design review (PDR) which is required before entering 
the detailed design phase (Phase C). In the PDR, the validity of various achievements in the 
basic design phase and the technical feasibility were examined. The review items included 
the effectiveness of the mission requirements at the time of the PDR and the validation 
of the development and verification plans. It was confirmed that the project is designed 
to ensure appropriate reliability to fulfill the mission goals as are listed in ``the mission 
system requirement'' document and that the project is feasible both technically and 
schedule-wise at the system, sub-systems, and component levels.  Based on the results 
of the PDR, ASTRO-H is now in Phase C since May 2010.

In this paper, we will summarize the scientific requirements, the mission concept and the current
baseline configuration of instruments
of ASTRO-H.

\begin{table}
\caption{ASTRO-H Mission}
\begin{center}
\begin{tabular}{ll}
\hline
 Launch site & 
Tanegashima Space Center, Japan\\
 Launch vehicle& JAXA H-IIA rocket\\
 Orbit Altitude& $\sim$550 km\\
 Orbit Type& Approximate circular orbit\\
 Orbit Inclination& $<$ 31 degrees\\
 Orbit Period& 96 minutes\\
Total Length& 14 m\\
 Mass& $<$ 2.6  metric ton\\
 Power& $<$ 3500 W\\
 Telemetry Rate& 8 Mbps (X-band QPSK) \\
 Recording Capacity& 12 Gbits at EOL\\
 Mission life & $>$ 3 years\\
\hline
 \end{tabular}
 \end{center}
\end{table}

\section{Science Requirements}

ASTRO-H aims to achieve the scientific goals and objectives listed below by inheriting 
the tradition and advancing the technology of  highly successful X-ray missions 
initiated by SAS beginning 
with the launch of the Hakucho mission in 1979 through to the currently operating Suzaku mission.

\begin{center}
\bf{Scientific goals and Objectives}
\end{center}

\subsubsection*{Revealing the large-scale structure of the Universe and its evolution}

\begin{itemize}
\item ASTRO-H will observe clusters of galaxies, the largest bound 
structures in the Universe, with an aim to reveal the interplay between 
the thermal energy of the intracluster medium, the kinetic energy 
of sub-clusters from which clusters form, measure the non-thermal energy;
and to directly trace the dynamic evolution of clusters of galaxies.

\item ASTRO-H will observe distant supermassive black holes hidden 
by thick intervening material  with  100 times higher sensitivity 
than Suzaku, and will study their evolution and  role in galaxy formation.
\end{itemize}

\subsubsection*{Understanding the extreme conditions in the Universe}

\begin{itemize}
\item ASTRO-H will measure the motion of matter very close to black 
holes with an aim to sense the gravitational distortion of  space, and to understand 
the structure of relativistic space-time.
\end{itemize}

\subsubsection*{Exploring the diverse phenomena of non-thermal Universe}
\begin{itemize}
\item ASTRO-H will derive the physical conditions of the sites where 
high energy particles (cosmic rays) gain energy and will elucidate 
the process in which gravity, collisions, and stellar explosions energize  
those cosmic rays.

\end{itemize}

\subsubsection*{Elucidating dark matter and dark energy}
\begin{itemize}
\item ASTRO-H will map the distribution of dark matter in clusters 
of galaxies and will determine the total mass of galaxy clusters 
at different distances (and thus at different ages), and will study the role 
of dark matter and dark energy in the evolution of these systems.

\end{itemize}

In order to perform the leading edge science described above,
ASTRO-H is designed with cutting edge technology.
With an unprecedented spectroscopic capability and a wide-band energy coverage, ASTRO-H
will measure the motion of hot gas, depicting the dynamic nature of the 
evolution of the Universe.  These measurements will be the key in a great pursuit 
to understand the origin of the dark matter filling in the Universe.

\section{Spacecraft and Instruments}

\begin{figure}[tb]
\centerline{\includegraphics[width=17cm]{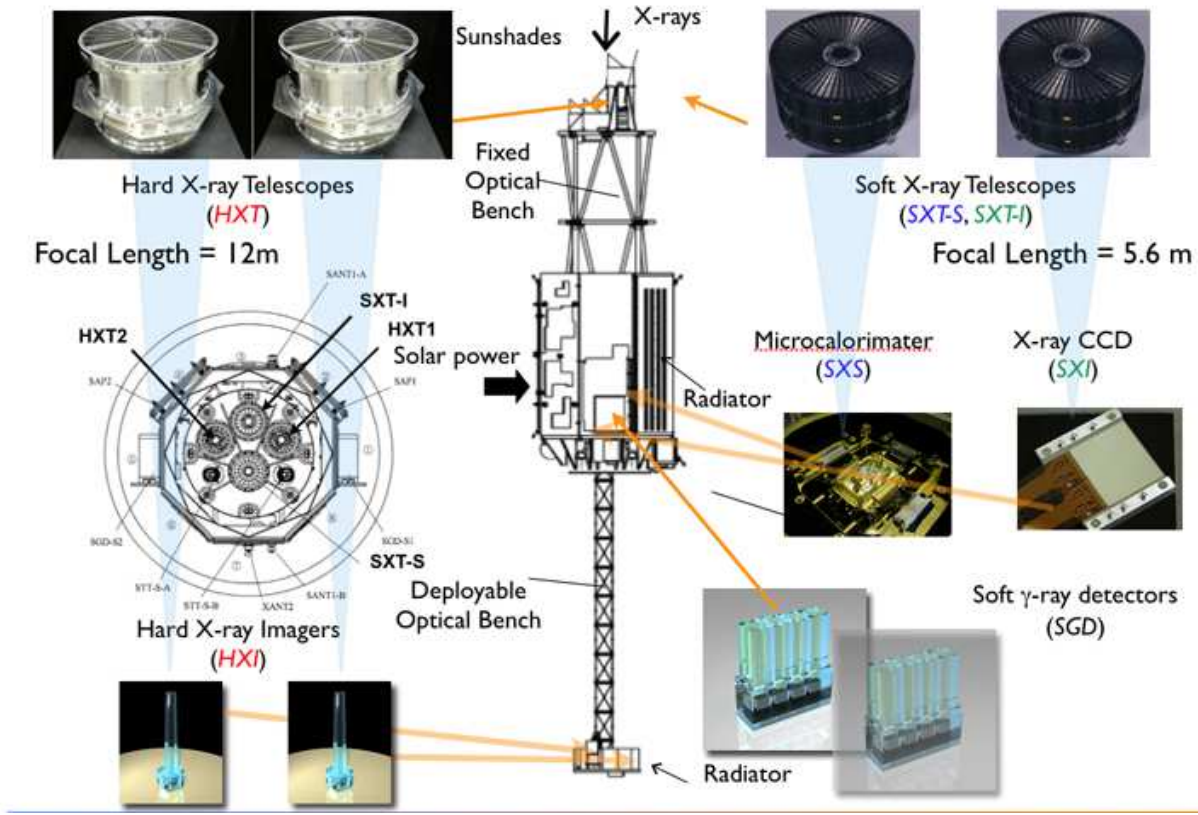}}
\caption{Configuration of the ASTRO-H satellite.}
\label{Fig:Config}
\end{figure}
 
ASTRO-H will carry two Hard X-ray Telescopes (HXTs) for the Hard X-ray Imager (HXI), 
and two Soft X-ray Telescopes (SXTs), one with a micro-calorimeter spectrometer 
array  with excellent energy resolution of $\sim$7\,eV, and the other with a 
large area CCD in their respective focal planes.  In order to extend the energy 
coverage to the soft $\gamma$-ray region up to 600 keV, the Soft Gamma-ray 
Detector (SGD) will be implemented as a non-focusing detector.  With these instruments, 
ASTRO-H will cover the bandpass between 0.3~keV and 600~keV.  The conceptual design 
of each instrument is shown in Fig.~\ref{Fig:Config} and the requirements and specifications of 
those instruments based on the base-line design are 
summarized in Table.~\ref{Table:Spec} and ~\ref{Table:Spec2}.
Detailed descriptions of the instruments are available  in other papers in these  proceedings
\cite{Ref:Kunieda2010,Ref:Selemistos2010,Ref:Mitsuda2010,Ref:Kokubun2010,Ref:Tsunemi2010,Ref:Tajima2010}.

Both soft and hard X-ray mirrors are mounted on top of the fixed
optical bench. The two focal-plane detectors for soft X-ray 
mirrors are mounted on the base plate of the spacecraft,
while two hard X-ray detectors are mounted on the extensible 
optical bench to attain 12~m focal length.

\begin{table}
 
 \caption{Requirements to the instruments onboard ASTRO-H}
\begin{center}
  \label{Table:Spec}
 \begin{tabular}{|p{5cm}|p{2.6cm}|p{7.6cm}|}
\hline
Science Payload  & Energy Coverage & Requirements \\
\hline
 Hard X-ray Imaging System : 
(HXT+HXI) &
5--80 keV & Sensitivity for detecting point sources with a brightness of 1/100,000 times fainter than the Crab nebula at $>$10 keV.  Energy resolution to resolve background emission lines by activated material.\\
\hline
Soft X-ray Spectrometer System:
 (SXT-S+XCS) &
0.3--12 keV&
Energy resolution corresponding to 300 km/s in the Doppler width of iron emission lines at 6 keV for representative clusters of galaxies. Fields of view covering 1/5 of the virial radius for representative clusters at a redshift of 0.1.\\
\hline
Soft X-ray Imaging System (SXT-I+SXI) &
0.4--12 keV &
Fields of view twice as large as those of the Hard X-ray Imaging System and the Soft X-ray Spectrometer System for obtaining accurate wide energy range spectra and observing surrounding area of the main targets.\\
\hline
Soft $\gamma$-ray detector (SGD)&
40--600 keV&
Sensitivity for obtaining spectra up to 600 keV for more than 10 point sources (spectral index of 1.7) with a brightness of 1/1,000 times fainter than the Crab nebula.
\\
\hline
 \end{tabular}
 \end{center}
 \end{table}

ASTRO-H is in many ways similar to Suzaku in terms of orbit,
pointing, and tracking capabilities, although the mass is considerably larger; the total mass 
at launch will be 2600~kg  (compared with Suzaku's 1700 kg).
 ASTRO-H will be launched into a circular orbit 
with altitude 500--600~km, and inclination 31~degrees or less.  Science 
operations will be similar to those of Suzaku, with pointed observation of each 
target until the integrated observing time is accumulated, and then slewing 
to the next target. A typical observation will require 40--100~ksec integrated 
observing time, which corresponds to 1--2.5~days of clock time. All instruments  are co-aligned and 
will operate simultaneously.

 \begin{table}
\caption{Specification of instruments}
\label{Table:Spec2}
\begin{small}
\begin{center}
\hspace{5mm}
\begin{tabular}[htp]{l|l|ll}
\hline \hline
Hard X-ray Imaging System &\multicolumn{3}{l}{HXT (Hard X-ray Telescope)/HXI (Hard X-ray Imager)} \\ \cline{2-4}
            & &Focal Length & 12 m\\
               & &Effective Area &300 cm$^{2}$ (at 30 keV) \\
      & &Energy Range&5--80 keV \\
              & &Angular Resolution & $<$1.7 arcmin (HPD)  \\
              & &Effective FOV & $\sim$9 $\times$ 9 arcmin (12 m Focal Length)  \\           
 & & Energy Resolution & $<$ 1.5 keV (FWHM, at 60 keV)\\
& & Timing Resolution & several 10 $\mu$s \\
& & Detector Background & $<$ 1--$3 \times 10^{-4}$ cts s$^{-1}$ cm$^{-2}$ keV$^{-1}$ \\
& & Operating Temperature & $\sim$ $-20$ $^\circ$C \\
\hline \hline
Soft X-ray Spectrometer System&\multicolumn{3}{l}{SXT-S (Soft X-ray Telescope)/ XCS (X-ray Calorimeter Spectrometer) }  \\ \cline{2-4}
           & &Focal Length & 5.6 m\\
  & &Effective Area & 210 cm$^{2}$  (at 6 keV) \\
  & && 160 cm$^{2}$  (at 1 keV) \\
   & &Energy Range & 0.3--12 keV \\
& &Angular Resolution & $<$ 1.7 arcmin (HPD)  \\
 & &Effective FOV & $\sim$ 3 $\times$ 3 arcmin  \\         
& &Energy Resolution&  7 eV (FWHM, at 6 keV)\\
& & Timing Resolution & 80 $\mu$s \\
  & & Detector Background &  $<$$6  \times 10^{-3}$ cts s$^{-1}$ cm$^{-2}$ keV$^{-1}$\\
& & Operating Temperature & 50 mK \\
\hline \hline
Soft X-ray Imaging System &\multicolumn{3}{l}{SXT-I (Soft X-ray Telescope)/SXI (Soft X-ray Imager)} \\ \cline{2-4}
           & &Focal Length & 5.6 m\\
  & &Effective Area & 360 cm$^{2}$  (at 6 keV) \\
               & &Energy Range &  0.4--12 keV  \\
            & &Angular Resolution & $<$ 1.7 arcmin (HPD)  \\
               & &Effective FOV & $\sim$ 38 $\times$ 38 arcmin   \\      
 & &  Energy Resolution & $<$ 150 eV (FWHM, at 6 keV) \\
 & & Timing Resolution & 4 sec \\
  & & Detector Background &  $<$ a few $\times 10^{-3}$ cts s$^{-1}$ cm$^{-2}$ keV$^{-1}$\\
& & Operating Temperature & $-120$ $^\circ$C\\

\hline \hline
Soft Gamma-ray non-Imaging System& \multicolumn{3}{l}{SGD (Soft Gamma-ray Detector)}   \\ \cline{2-4}
    & &Energy Range & a few 10 keV$-$600 keV  \\
   & &Energy Resolution & $\sim$ 2 keV (FWHM, at 40 keV) \\
   & &Geometrical Area  &   210 cm$^{2}$  \\
   & &Effective Area&   $\sim$ 30 cm$^{2}$ (Compton mode, at 100 keV) \\
   & &FOV  & 0.6 $\times$ 0.6 deg$^{2}$  ($<$ 150 keV) \\
 & & Timing Resolution & several 10 $\mu$s \\
 & & Detector Background  & $<$ a few $\times 10^{-6}$ cts s$^{-1}$ cm$^{-2}$ keV$^{-1}$ \\
 & & & ( 100--200 keV) \\
& & Operating Temperature & $\sim$ $-20$ $^\circ$C\\
\hline \hline
\end{tabular}
\end{center}
\end{small}
\end{table}

 \subsection{Hard X-ray Imaging System}

\begin{figure}[htb]
\centerline{\includegraphics[width=12cm]{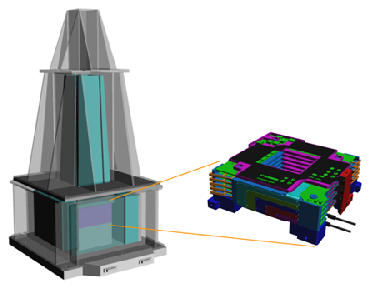}}
\caption{The Hard X-ray Imager. A stack of Si and CdTe double sided cross-strip detectors is
mounted in a well-type BGO shield.
}
\label{Fig:HXI}
\end{figure}

The hard X-ray imaging system onboard ASTRO-H consists of two identical mirror-detector 
pairs (HXT and  HXI).
The HXT has conical-foil mirrors with depth-graded multilayer reflecting surfaces 
that provide a 5--80~keV energy range\cite{Ref:Ogasaka,Ref:Kunieda2010}.   The effective area of the HXT
is maximized for a long focal length, with current design value of 12~m
giving an effective area of $\sim$300~cm$^{2}$ at 30~keV for two HXTs. 
A depth-graded multilayer mirror reflects X-rays not only by total
external reflection but also by Bragg reflection. In order to obtain
high reflectivity up to 80~keV, the HXTs consist of a stack of
multilayers with different sets of periodic length and number of layer
pairs with a platinum/carbon coating. The technology of a hard X-ray
focusing mirror has already been proved by the balloon programs
InFOC$\mu$S (2001, 2004)\cite{Ref:Kunieda2006,Ref:Ogasaka:In}, 
HEFT (2004)\cite{Ref:Fiona} and SUMIT (2006)\cite{Ref:Kunieda2006}.

The non-imaging instruments flown so far were essentially limited to
studies of sources with 10--100~keV fluxes of at best $>$4 $\times$ 10$^{-12}$--%
10$^{-11}$~erg~cm$^{-2}$s$^{-1}$. This limitation is due to the presence
of high un-rejected backgrounds from particle events and Cosmic X-ray
radiation, which increasingly dominate above 10~keV. Imaging, and
especially focusing instruments have two tremendous advantages. Firstly,
the volume of the focal plane detector can be made much smaller than for
non-focusing instruments, thus reducing the absolute background level
since the background flux generally scales with the size of the
detector.  Secondly, the residual background, often time-variable, can
be measured simultaneously with the source, and can be reliably
subtracted.  For these reasons, a focusing hard X-ray telescope in
conjunction with an imaging detector sensitive for hard
X-ray photons is the appropriate choice to achieve a breakthrough in
sensitivity for the field of high energy astronomy. In addition to the improvement
of sensitivity, the HXI provides a ``true" imaging capability which enable us to
study spatial distributions of hard X-ray emission.

The HXI consists of four-layers of 0.5~mm thick Double-sided Silicon Strip
Detectors (DSSD) and one layer of 0.75~mm thick CdTe imaging 
detector (Fig.~\ref{Fig:HXI})\cite{Ref:Takahashi_SPIE1,Ref:Nakazawa,Ref:Kokubun,Ref:Kokubun2010}.  
In this configuration, soft X-ray photons below $< \sim$ 20 keV are 
absorbed in the Si part (DSSD), while hard X-ray photons above  $\sim$ 20 keV go through the
Si part and are detected by the newly developed CdTe double sided cross-strip detector. 
The low energy spectrum, obtained with Si, is less contaminated by 
the background due to activation in heavy material, such as Cd and Te.  Fast timing
response of silicon strip detector and CdTe strip detector allows us to
place the entire detector inside a very deep well of the active shield
made of BGO (Bi$_{4}$Ge$_{3}$O$_{12}$) scintillators. Signal from the BGO
shield is used to reject background events. 
The DSSDs cover the energy below 30~keV while the 
CdTe strip detector covers the 20--80~keV band.
Each DSSD has a size of 3.2$\times$3.2~cm$^{2}$
and a thickness of 0.5~mm, resulting in 2~mm in total thickness, the same as that of the PIN detector 
of the HXD onboard Suzaku. A CdTe strip
detector has a size of 3.2$\times$3.2~cm$^{2}$ and a thickness of 0.75~mm. 
In addition to the increase of efficiency, the stack configuration and
individual readout provide information on the interaction depth. This
depth information is very useful to reduce the background in space
applications, because we can expect that low energy X-rays interact in
the upper layers and, therefore, it is possible to reject the low energy events
detected in lower layers. Moreover, since the background rate scales
with the detector volume, low energy events collected from the first few
layers in the stacked detector have a high signal to background ratio,
in comparison with events obtained from a monolithic detector with a
thickness equal to the sum of all layers.  

\subsection{Soft X-ray Spectrometer System}


The soft X-ray Spectrometer (SXS) consists of the Soft X-ray Telescope 
(SXT-S), the X-ray Calorimeter Spectrometer (XCS) and the cooling system\cite{Ref:Mitsuda,Ref:Mitsuda2010}.
The XCS is a 36~pixel system with an energy resolution of $\leq$7~eV between 0.3--12~keV. 
Micromachined, ion-implanted silicon is the basis of the thermistor array, 
and 8-micron-thick HgTe absorbers provide high quantum efficiency across the 
0.3--12~keV band. With a 5.6-m focal length, the 0.83~mm pixel pitch corresponds to 
0.51~arcmin, giving the array a field of view of 3.04~arcmin on a side.  
The detector assembly provides electrical, thermal, and mechanical interfaces 
between the detectors (calorimeter array and anti-coincidence particle detector) 
and the rest of the instrument.  Soft X-ray Telescope (SXT) for XCS
is an upgraded version of the Suzaku X-ray telescope with an improved
angular resolution and a larger area\cite{Ref:Okajima,Ref:Selemistos2010}.

\begin{figure}
\centerline{{\includegraphics[height=8cm,clip]{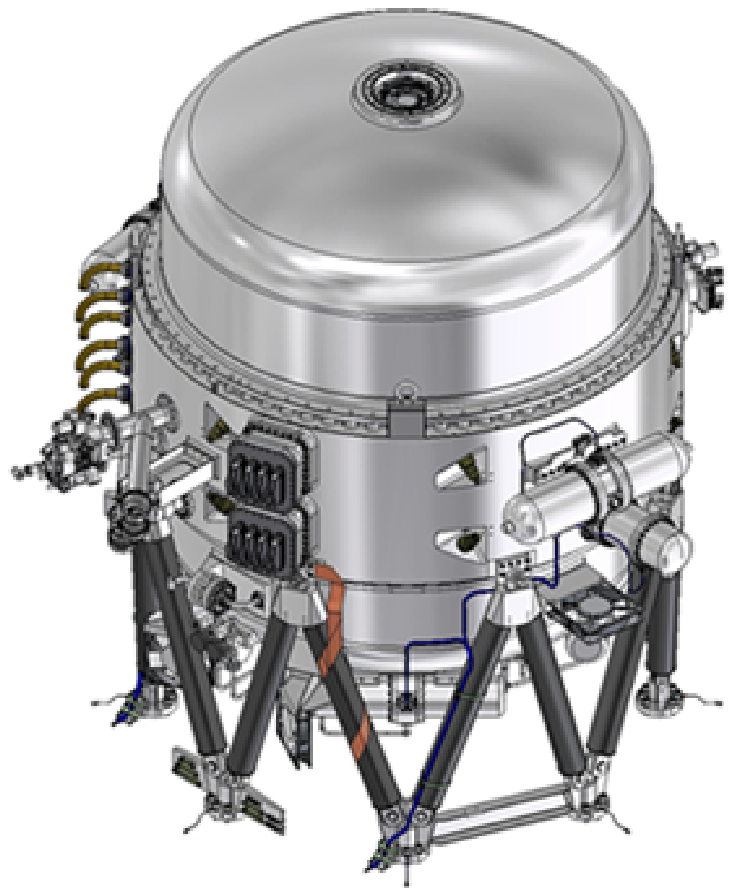}}{\includegraphics[height=8cm,clip]{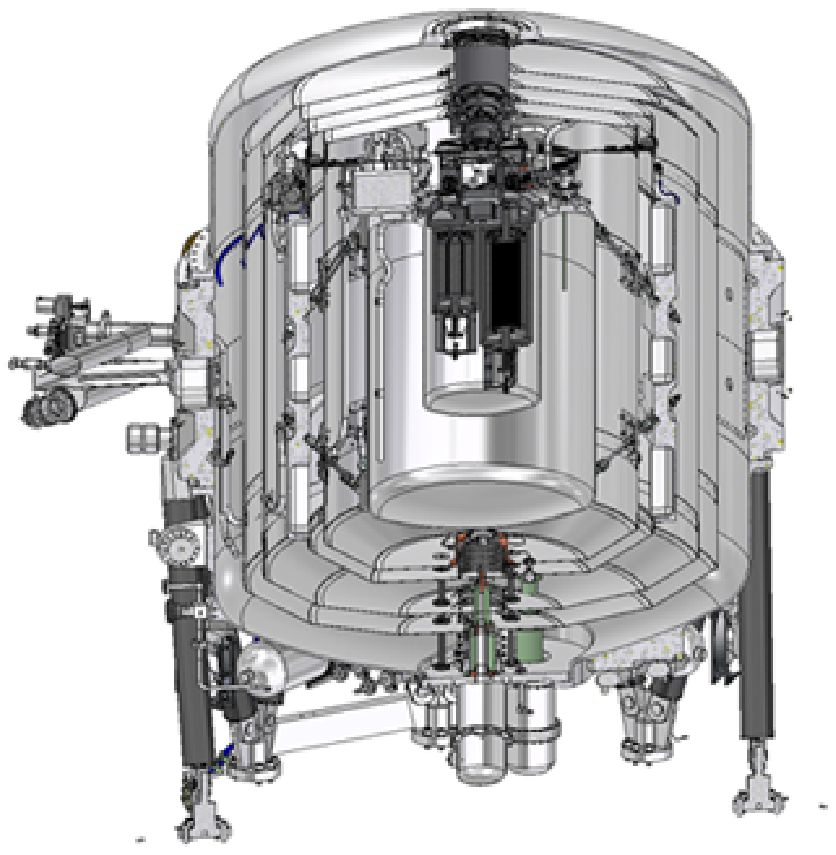}}}
\caption{Outlook   and cross sectional view of the SXS dewar. The outer shell of the dewar is 950 mm in diameter. }
\end{figure}

The SXS science objectives require a mirror with larger effective area than 
those flown on Suzaku, especially in the Fe K band. This is attained by the 
combination of increased focal length and larger diameter. The SXS effective 
area at 6~keV is 210~cm$^{2}$, a 60~\% increase over the Suzaku XRS,  while at 1~keV 
the SXS has 160~cm$^{2}$, a 20~\% increase.  The required angular resolution 
is 1.7~arcmin, HPD, comparable to the in-orbit performance of the mirrors on Suzaku.  

The XCS cooling system must cool the array to 50~mK with sufficient duty cycle to 
fulfill the SXS scientific objectives:  this requires extremely low heat loads.  
To achieve the necessary gain stability and energy resolution, the cooling 
system must regulate the detector temperature to within 2~$\mu$K rms for at 
least 24~hours per cycle.  From the detector stage to room temperature, the 
cooling chain is composed of a 3-stage  Adiabatic Demagnetization 
Refrigerator (ADR), superfluid liquid $^{4}$He (hereafter LHe), 
a $^{4}$He Joule-Thomson (JT) cryocooler, and 2-stage Stirling cryocoolers.  
As with Suzaku, the array will be cooled using an ADR\cite{Ref:Fujimoto2010,Ref:Shirron}.  
An ADR has been adopted because it readily meets 
the requirements for detector temperature, stability, recycle time, 
reliability in a space environment, and previous flight heritage.  The 
design of Stirling cryocoolers is based on coolers developed for space-flight 
missions in Japan (Suzaku, AKARI, and the SMILES instrument  deployed on 
the ISS\cite{Ref:Narasaki}) that have achieved excellent performance with 
respect to cooling power, efficiency and mass. As a heat-sink for the 2-stage 
ADR, 30 L of LHe is used. To reduce the parasitic heat load on the He tank, a 
$^{4}$He JT cryocooler is used to cool a 4 K shield. To achieve redundancy for failure 
(unexpected loss) of LHe, another  ADR (3rd stage ADR) is used between the 
He tank and the JT cryocooler, with two heat-switches on both sides.  
This ADR is operated if LHe is lost, to cool down the 1 K shield (He tank). 
A series of five blocking filters shield the calorimeter array from UV and 
longer wavelength radiation. The aluminized polyimide filters are similar 
to those successfully flown on Suzaku.

\begin{figure}
\centerline{\includegraphics[height=7.5cm,clip]{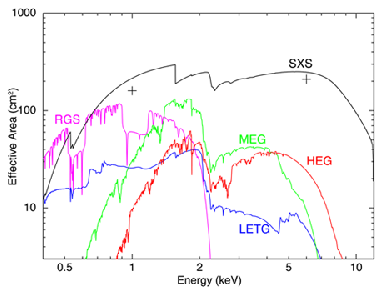}
                   {\includegraphics[height=5.8cm,clip]{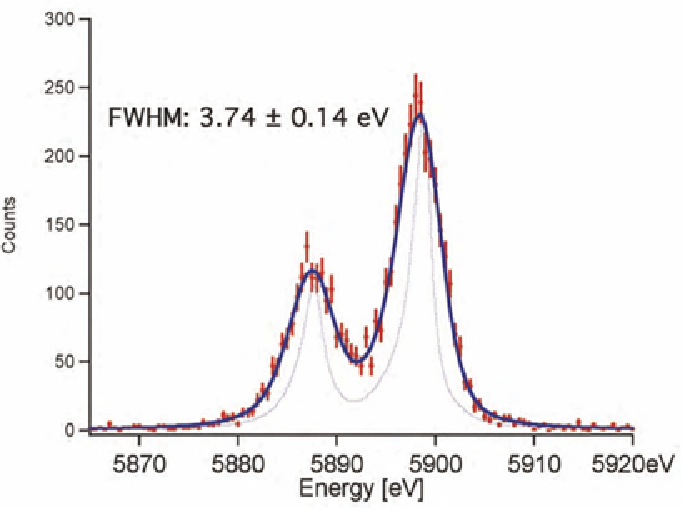}}}
\caption{(left) The effective area of SXS (right). The energy resolution obtained from Mn K$\alpha_{1}$using a  detector from the XRS program but with a new sample of absorber material (HgTe) that has lower specific heat, leading to an energy resolution of 3.7~eV (FWHM).  The SXS could have an energy resolution approaching this value \cite{Ref:Porter2,Ref:Kelley,Ref:Porter}.}
\label{Fig:XRS}
\end{figure}

In combination with a high throughput X-ray telescope, the 
SXS improves on the Chandra and XMM-Newton grating spectrometers in 
two important ways. At E $>$ 2~keV, SXS is both more sensitive and has 
higher resolution (Fig.\ref{Fig:XRS}), especially in the Fe K band where 
SXS has 10 times the collecting area and much better energy resolution, giving a net 
improvement in sensitivity by a factor of 30 over Chandra.  The broad bandpass of 
the SXS encompasses the critical inner-shell emission and absorption lines 
of Fe I-XXVI between 6.4 and 9.1~keV. Fe K lines provide particularly useful 
diagnostics because of their (1) strength, due to the high abundance and large 
fluorescent yield (30\%), (2) spectral isolation from other lines, and (3) 
relative simplicity of the atomic physics. Fe K emission lines reveal conditions 
in plasmas with temperatures between 10$^{7}$ and 10$^{8}$ K, which are typical values for 
stellar accretion disks, SNRs, clusters of galaxies, and many stellar coronae.  
In cooler plasmas, Si, S, and Fe fluorescence and recombination occurs 
when an X-ray source illuminates nearby neutral material. Fe emission 
lines provide powerful diagnostics of non-equilibrium ionization 
due to inner shell K-shell transitions from Fe XVII--XXIV\cite{Ref:Decaux}.

In order to obtain a good performance for bright sources,
a filter wheel (FW) assembly, which includes a wheel with selectable filters and a set of modulated X-ray sources 
will be used at a distance of 90 cm from the detector. The FW is able to rotate a suitable filter into the
beam to optimize the quality of the data, depending on the source characteristics\cite{Ref:Vries}. In addition to the
filters, a set of on-off-switchable X-ray calibration sources, using light sensitive photo-cathode, will be implemented.
These calibration sources will allow proper gain and linearity calibration of the detector in flight.

SXS uniquely performs high-resolution spectroscopy of extended sources. 
In contrast to a grating, the spectral resolution of the calorimeter is 
unaffected by source's angular size because it is non-dispersive.  
For sources with angular extent larger than 10~arcsec, Chandra MEG 
energy resolution is degraded compared with that of a CCD; the energy 
resolution of the XMM-Newton RGS is similarly degraded for sources 
with angular extent $\geq$2~arcmin. SXS makes possible high-resolution 
spectroscopy of sources inaccessible to current grating instruments.

The key properties of SXS are its high spectral resolution for both 
point and diffuse sources over a broad bandpass ($\leq$7~eV FWHM throughout 
the 0.3--12~keV band), high sensitivity (effective area of 160~cm$^2$ at 1~keV 
and 210~cm$^2$ at 7~keV), and low non-X-ray background 
(1.5$\times$10$^{-3}$~cts~s$^{-1}$keV$^{-1}$). These properties open up 
a full range of plasma diagnostics and kinematic studies of X-ray emitting 
gas for thousands of targets, both Galactic and extragalactic. SXS improves 
upon and complements the current generation of X-ray missions, including 
Chandra, XMM-Newton, Suzaku and Swift. 

\subsection{Soft X-ray Imaging System}
\begin{figure}
\centerline{\includegraphics[height=9cm,clip]{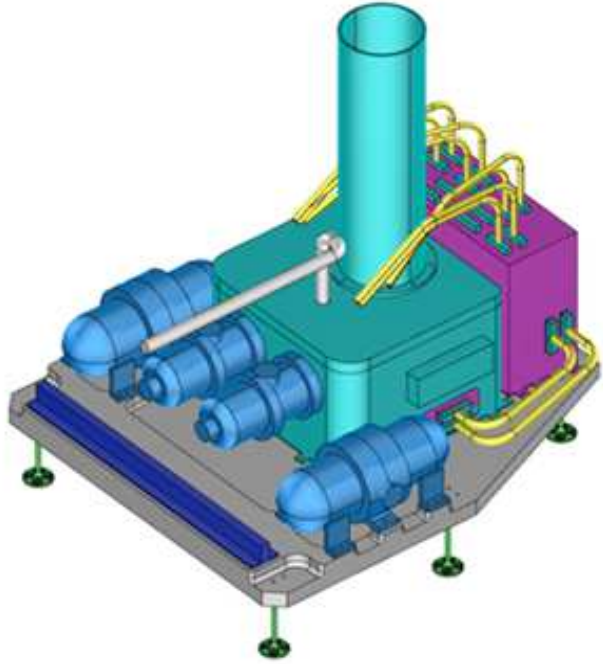}}
\caption{Schematic drawing of the Soft X-ray Imager (SXI and a picture of a prototype CCD chip). }
\label{SXI-OUTLOOK}
\end{figure}

X-ray sensitive silicon charge-coupled devices (CCDs) are a key device
for the X-ray astronomy. The low background and high energy resolution
achieved with the XIS/Suzaku clearly show that the X-ray CCD will also
play very important role in the ASTRO-H mission.  Soft X-ray imaging system 
will consist of an imaging mirror and a CCD camera (Soft X-ray Telescope (SXT-I) 
and Soft X-ray Imager (SXI))\cite{Ref:Tsuru,Ref:Tsunemi,Ref:Tsunemi2010}. Fig.~\ref{SXI-OUTLOOK} 
 shows an schematic drawing of
SXI.
 
In order to cover the soft X-ray band below 12~keV,
the SXI will use next generation Hamamatsu CCD chips with
a thick depletion layer of 200 $\mu$m, low noise, and almost no cosmetic defects. 
The SXI features a large FOV and covers a 38$\times$38~arcmin$^{2}$ region on the sky,
complementing the smaller FOV of the 
SXS calorimeter. A mechanical cooler ensures a long operational life at 
$-$120~$^\circ$C\@. The overall quantum efficiency and spectral resolution 
is better than the Suzaku XIS\@.  The imaging mirror has a 5.6-m focal 
length, and a diameter of 45~cm.

\subsection{Soft Gamma-ray Detector (SGD)}

Highly sensitive observations in the energy range above the HXT/HXI
bandpass are crucial to study the spectrum of X-rays arising from accelerated particles. 
The SGD is a non-focusing soft gamma-ray detector with a
40--600~keV energy range and sensitivity at 300~keV of more than
10 times better than the Suzaku HXD (Hard X-ray Detector).
It outperforms previous soft-$\gamma$-ray instruments in background
rejection capability by adopting a new concept of narrow-FOV Compton
telescope\cite{Ref:Takahashi_Yokohama,Ref:Takahashi_SPIE2}.

In order to lower the background dramatically and thus to improve the
sensitivity as compared to the HXD of Suzaku, the design combines a stack of Si
and CdTe pixel detectors to form a Compton camera.  The
telescope is then mounted inside the bottom of a well-type active
shield.  Above $\sim$40~keV, each valid event is required to interact twice in the stacked
detector, once by Compton scattering in a stack of Si strip detectors,
and then by photo-absorption in the CdTe part (Compton mode).  Once the
locations and energies of the two interactions are measured, the Compton
kinematics allows the calculation of the energy and direction (as a cone in
the sky) of the incident $\gamma$-ray.

\begin{figure}
\centerline{\includegraphics[height=7.0cm,clip]{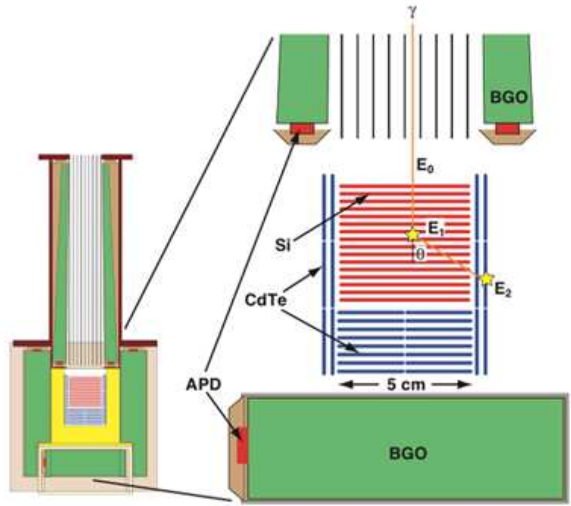}}
\caption{Conceptual drawing of an SGD Compton camera unit.}
\label{Fig:SGD_Concept}
\end{figure}

As shown schematically in Fig.~\ref{Fig:SGD_Concept}, the telescope
consists of 32 layers of 0.6~mm thick Si pad detectors 
and eight layers of CdTe pixellated detectors with a thickness of 0.75~mm.  The sides are 
also surrounded by two layers of CdTe pixel detectors. The 
opening angle provided 
by the BGO shield is $\sim$10~degrees at 500~keV. As 
compared to the HXD, the shield part is made compact by adopting the 
newly developed avalanche photo-diode.  
An additional PCuSn collimator restricts the field of view of the
telescope to 30' for photons below 100 keV  to minimize the
flux due to the Cosmic X-ray Background in  the FOV.  These modules
are then arrayed to provide the required area. 

ASTRO-H will have two SGD detectors, each consisting of four units. Each
detector will be mounted separately on two sides of the satellite.
It should be noted that when the Compton condition is not used (Photo absorption mode),
the stacked DSSD  can be used as a standard photo-absorption type 
detector with  the total thickness $\sim$20~mm of silicon. The detector then
covers energies above 10~keV as a collimated-type $\gamma$-ray detector.
The effective area of the SGD is $>$30~cm$^{2}$ at 100~keV in the
Compton mode.
Since the  scattering angle of gamma-rays can be measured
via reconstruction of the Compton scattering in the Compton camera, the
SGD is capable of measuring polarization of celestial sources brighter
than a few $\times$ 1/100 of the Crab Nebula, polarized 
above $\sim$ 10 \%. This capability is expected to yield polarization measurements 
in several celestial objects, providing new insights into properties of soft gamma-ray
emission processes\cite{Ref:Tajima-pol,Ref:Tajima2010}.

\section{Expected Scientific Performance}

The ASTRO-H mission objectives are: to study the evolution of yet-unknown 
obscured supermassive black holes (SMBHs) in Active Galactic Nuclei (AGN); 
to trace the growth history of the largest structures in the Universe;
to provide insights into the behavior of material in extreme gravitational fields; 
to determine the spin of black holes and the equation of state of neutron stars; 
to trace particle acceleration structures in clusters of galaxies and SNRs; 
and to investigate the detailed physics of astrophysical jets. 

\begin{figure}[htbp]
\includegraphics*[width=8cm]{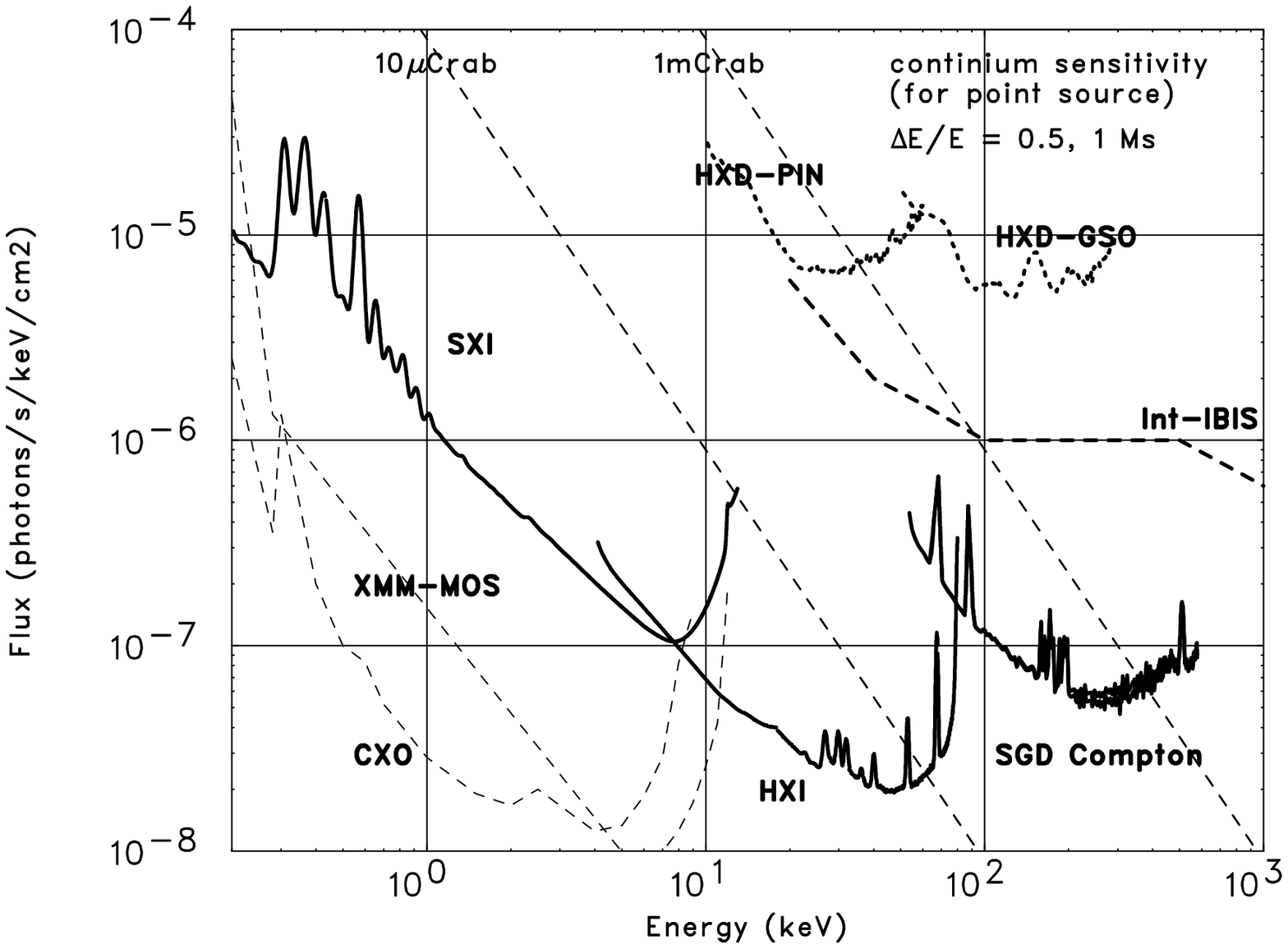}
\includegraphics*[width=8cm]{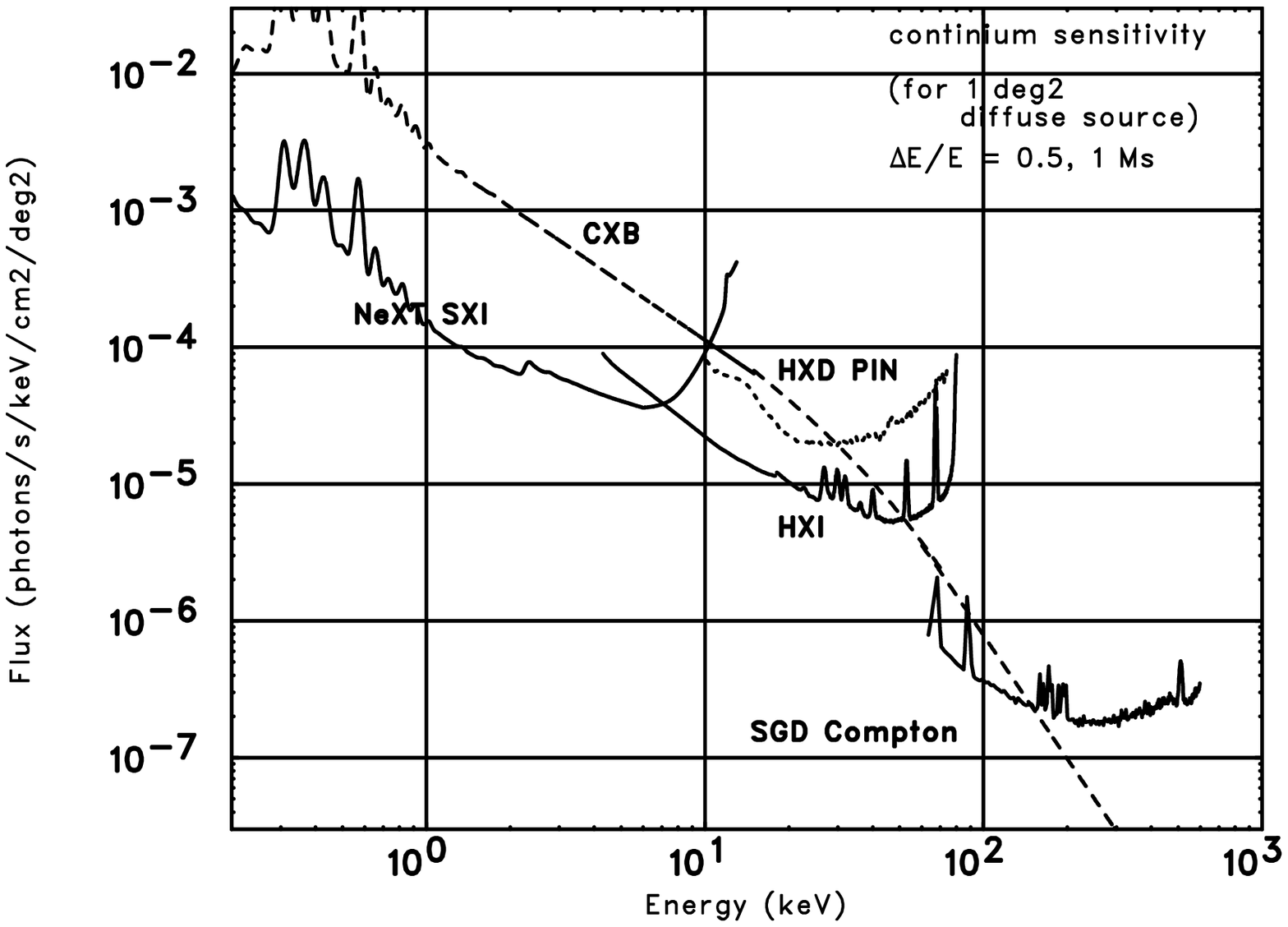}
\caption{Detection limits of the SXT-I/SXI, HXT/HXI and SGD for (a) point sources  and (b) for sources with of 1$\times$1~deg$^{2}$
 extension
(bottom) as functions of X-ray energy, where the spectral binning with ${\Delta}E/E = 0.5$ and 1000~ksec exposure are assumed.
Detection limits for the XMM-Newton, Chandra, Suzaku and Integral observatories are shown for comparison for the sensitivity for point sources.}
\label{fig:sensitivity-1}
\end{figure}

With ASTRO-H, we expect to achieve an area of about 300\,cm$^{2}$ at
30\,keV with a typical angular resolution of 1.7 arcmin (HPD). Fig.~\ref{fig:sensitivity-1}
shows  detection limits of the SXT-I/SXI, HXT/HXI and SGD for point sources and for 
sources with of $1  \times  1$ deg$^{2}$ extension.

The imaging capabilities at high X-ray energies open the completely
new field of spatial studies of non-thermal emission above 10~keV. This
will enable us to track  the evolution of active galaxies with accretion flows which are
heavily obscured, in order to accurately assess their contribution to
the cosmic X-ray background over cosmic time.   It will also uniquely allow
mapping of the spatial extent of the hard X-ray emission in diffuse sources, thus 
tracing the sites of particle acceleration in structures ranging in size from 
clusters of galaxies down to supernova 
remnants\cite{Ref:Koyama,Ref:Uchiyama,Ref:Aharonian}.  
Those studies will be complementary to the SXS measurements:  
observing the hard X-ray synchrotron emission will allow a study of the most energetic 
particles, thus revealing the details of particle acceleration mechanisms 
in supernova remnants, while the high resolution SXS data on the gas kinematics 
of the remnant will constrain the energy input into the accelerators.

SXS spectroscopy of extended sources can reveal line broadening and Doppler shifts 
due to turbulent or bulk velocities. This capability enables 
the spectral identification of cluster mergers, SNR ejecta dispersal patterns, the 
structure of AGN and starburst winds, and the spatially dependent abundance pattern 
in clusters and elliptical galaxies. SXS can also measure the optical depths of 
resonance absorption lines, from which the degree and spatial extent of turbulence 
can be inferred. Additionally, SXS can reveal the presence of relatively rare elements 
in SNRs and other sources through its high sensitivity to low equivalent width emission 
lines. The low SXS background ensures that the observations of almost all line rich 
objects will be photon limited rather than background limited. 

\subsection{Supernova Remnants}

The high resolution X-ray spectroscopy provided by ASTRO-H will be particularly 
ground-breaking for supernova remnants (SNRs) because they are extended objects 
with rich emission-line spectra from a wide range of different elements (carbon 
through nickel). To determine the element abundances reliably, measurements of 
the relative strengths of a number of lines from each elemental species are 
required. Accurate element abundances provide constraints to test the explosion 
mechanisms of supernovae and to explore their environments.  Gas motions of 
the rapidly expanding supernova ejecta and swept-up interstellar/circumstellar  
medium may also be measured by ASTRO-H via their Doppler shifts.  
Velocity measurements inferred from such Doppler shifts are needed to understand how SNRs evolve, 
based on their age and the detailed properties of the explosion, the ejecta, 
and ambient medium.

\begin{figure}
\centerline{\includegraphics[height=6cm,clip]{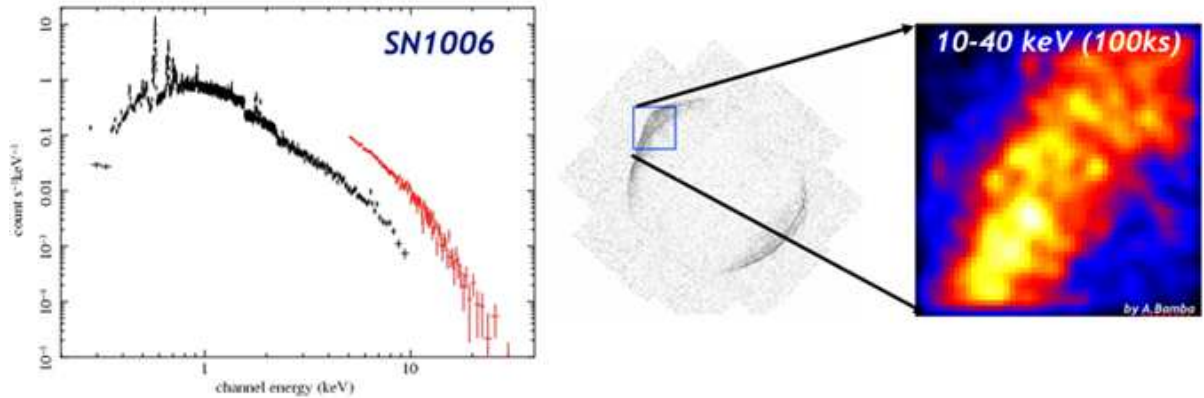}}
\caption{(left)  Simulated spectra for 100 ks SXS observation of SN1006 with SXS and HXI. (right) Image expected from the combination of HXT and HXI.}
\label{Fig:SNR}
\end{figure}

 Particle acceleration is receiving much attention at present, 
but the origin of cosmic rays is still unclear 100 years after their discovery.  
Young SNRs with shock speeds of several 1000 km/s are among the best candidates 
to accelerate cosmic rays up to the knee around $10^{15}$ eV (the highest energy 
accessible to Galactic accelerators). The combination of ASTRO-H's hard X-ray 
imaging capability and high spectral resolution will provide information to 
understand crucial aspects of shock acceleration in SNRs such as the maximum 
energy of the accelerated particles, the conditions at the acceleration sites, 
and the acceleration efficiency (See Fig.~\ref{Fig:SNR}).

 \subsection{A Census of Obscured Active Galactic Nuclei}

\begin{figure}
\centerline{\includegraphics[height=7cm,clip]{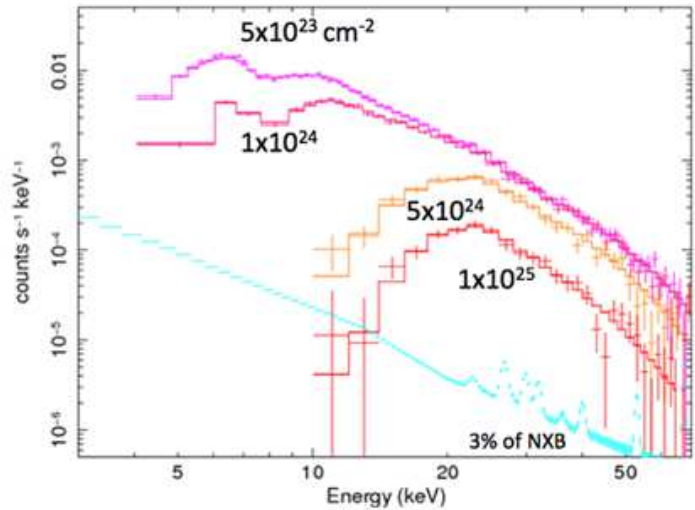}}
\caption{Simulated HXI spectra of heavily obscured AGN with different  absorbing columns (N$_{\rm{H}}$) for an exposure of 100 ks.  The continuum is assumed to be  a power-law of photon index 1.9, with an intrinsic 2-10 keV flux of 1x10$^{-11}$ erg/cm$^{2}/s$  (based on Swift J0601.9-8636\cite{Ref:Ueda2}). Scattered component and Fe lines are not included.}
\label{Fig:BH}
\end{figure}

Recent observations imply the existence of a large number of Active Galactic Nuclei (AGN) that are heavily obscured by the gas and dust surrounding their supermassive black holes\cite{Ref:Ueda2}. Some are identified as a "new type" of AGN,  so deeply buried in dense tori of gas that they show little emission in soft X-ray and visible light. While this has made such objects extremely difficult to detect and observe, this highly obscured activity may in fact represent the dominant phase of supermassive black hole growth.  Understanding this phase is thus key to  understanding  the correlated evolution of the black hole and its host galaxy. As shown in Fig.~\ref{Fig:BH}, the high sensitivity for hard X-rays provided by HXI allows  precise spectral studies of even very obscured AGN. ASTRO-H will provide us with a large AGN sample to pursue systematic  studies of the true AGN population, unbiased by obscuration effects, and to measure the co-evolution of supermassive black holes with their host galaxies.

By assuming a background level of $\sim$ 1$\times$10$^{-4}$ counts/s/cm$^{2}$/keV, in which
the non X-ray background is dominant, the source detection limit in
1000\,ksec in the $10- 80$ keV band would be roughly 10$^{-14}$\,erg\,cm$^{2}$\,s$^{-1}$ 
(for a power-law spectrum with a photon index of 2). This is about two orders 
of magnitude better than present instrumentation, and thus 
will result in a breakthrough in our understanding of
hard X-ray spectra. With this sensitivity, 40-50~\% of hard X-ray Cosmic
Background  would be be resolved\cite{Ref:Ueda} .
In addition to the imaging observations below 80~keV, the SGD will
provide a high sensitivity in the soft $\gamma$-ray region to match the
sensitivity of the HXT/HXI combination.  The extremely low background observations 
allowed by the new concept of a narrow-FOV Compton telescope adopted for
the SGD will provide sensitive $\gamma$-ray spectra up to 600~keV, with moderate
sensitivity for polarization measurements.  

\subsection{Close to Black Holes}

On the smallest scales, many active galactic nuclei (AGN) show signatures 
from the innermost accretion disk in the form of broad ``relativistic" Fe K 
emission lines. These broad lines were discovered using ASCA in the early 1990s 
and have been confirmed by XMM-Newton and Suzaku\cite{Ref:YTanaka,Ref:Reeves}. 
There is, however, a complex relationship between the Fe K line properties, the 
underlying continuum, and the signatures of cold and/or
partially ionized material near the AGN. Precise measurements of the complex 
Fe K line and absorption components require high spectral resolution.  
The measurement of changes in the X-ray emission and absorption spectral 
features on the orbital time  scale of black holes in AGN, will 
enable characterization of the velocity field and ionization state of the 
plasma closest to the event horizon.  The optically thick material that produces 
the broad fluorescent Fe K line also creates a Compton ``hump'' peaking 
at $E>20$~keV detectable with hard X-ray and soft gamma-ray detectors, 
providing multiple insights into the physics of the disk. In order to understand 
the evolution of environments surrounding supermassive black holes, 
high signal-to-noise measurements of the broad lines of many  AGN are needed 
up to, at least, $z$$\sim$2. This requires high spectral resolution and bandpass 
extending to at least $\sim$40~keV.  These observations will provide the first 
unbiased survey of broad Fe K line properties across all AGN.

XMM-Newton and Suzaku spectra frequently show time-variable absorption 
and emission features in the 5--10~keV band. If these features are due 
to Fe, they represent gas moving at very high velocities with both red-  
and blue-shifted components from material presumably near the event horizon. 
CCD resolution is insufficient, and the required grating exposures are 
too long to properly characterize the velocity field and ionization of 
this gas and determine whether it is from close to the black hole or 
from high velocity winds. SXS, in combination with  HXI, provides a 
dramatic increase in sensitivity over Suzaku, enabling measurements 
that probe the geometry of the central regions of $\sim$50 AGNs on 
the orbital timescale of the Fe producing region (for an AGN with 
a 3$\times$10$^{7}M_{\odot}$ black hole, this is  $\sim$60~$GM/c^{3}$ = 10~ksec). 

\subsection{Clusters of Galaxies}

\begin{figure}
\centerline{\includegraphics[height=7cm,clip]{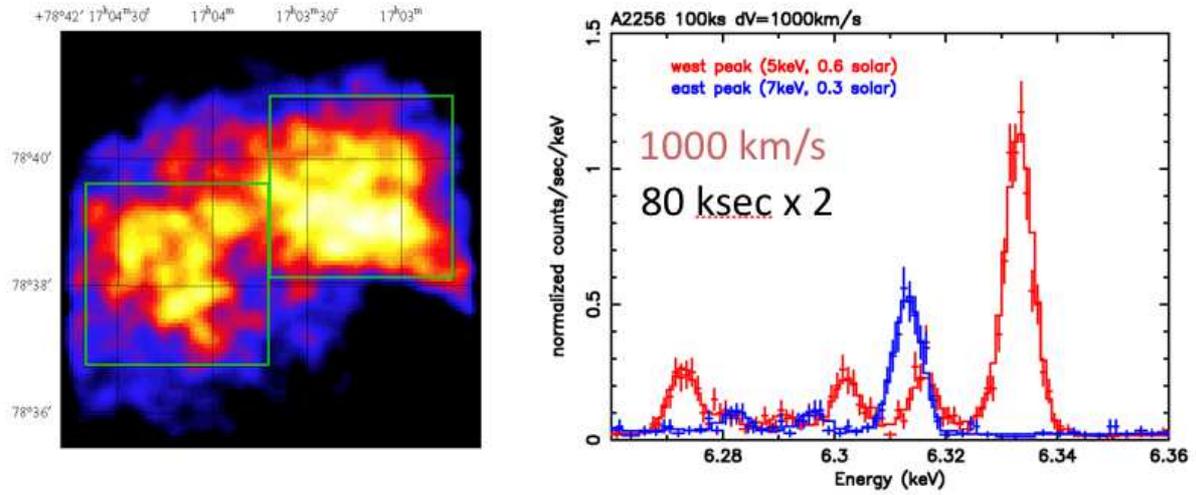}}
\caption{ Simulated image and spectra of a merging cluster A2256 by assuming
1000 km/s difference of line of sight speed of hot gas in the cluster. The simulation is for 80~ksec each.}
\label{Fig:SXS1}
\end{figure}

\begin{figure}
\centerline{\includegraphics[height=7cm,clip]{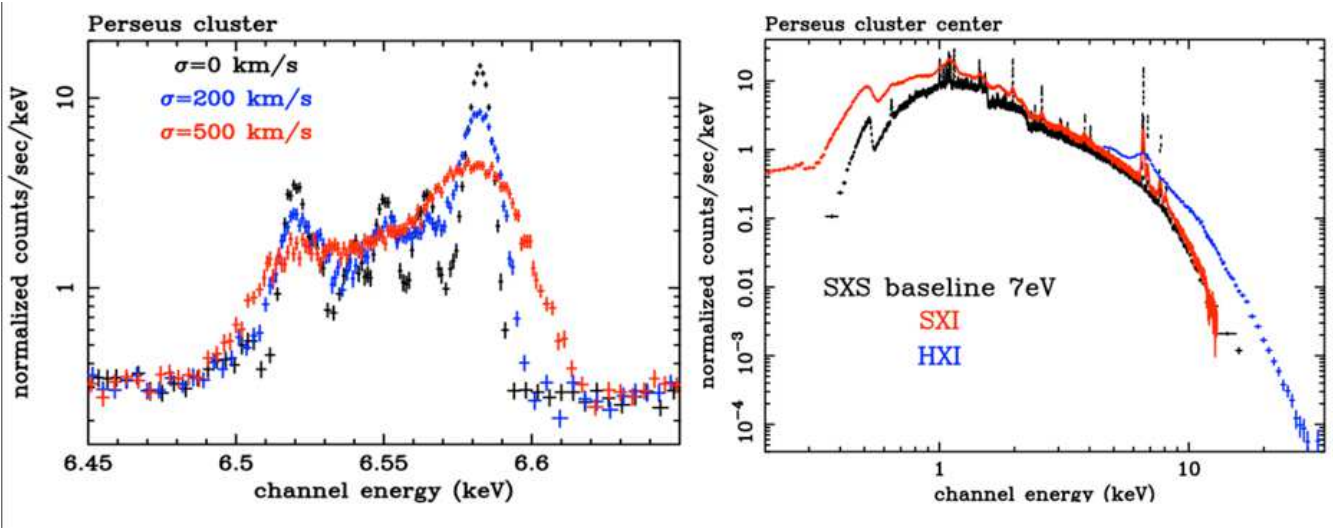}}
\caption{Simulated spectra for 100 ks SXS observation of Perseus Cluster. (left) Line profiles assuming $\sigma$=0, 100 and 200 km/s turbulence. (right) Wide band spectra obtained with SXS and HXI for hot plasma with three different temperature of 0.6 keV, 2.6 keV and 6.1 keV (r $<$ 2' )}
\label{Fig:SXS2}
\end{figure}

All studies of the total energy content (including that of non-thermal 
particles), aimed to draw a more complete picture of the
high energy universe, require observations by {\sl both} a spectrometer 
capable of measuring the bulk plasma velocities and/or turbulence with the 
resolution corresponding to the speed of a few $\times$ 100~km/s {\sl and}
an arc-min imaging system in the hard X-ray band, with the sensitivity two-orders of 
magnitude better than previous missions (See Fig.~\ref{Fig:SXS1} 
and Fig.~\ref{Fig:SXS2}).
In clusters, X-ray  emitting hot gas is trapped in the gravitational potential
well and shocks and/or turbulence are produced in this gas, as smaller
substructures with their own hot gas halos fall into and merge with the
dominant cluster. Large scale shocks can also be produced as gas from the
intracluster medium falls into the gravitational potential of a cluster.
Here, there is  a strong synergy between the hard 
X-ray imaging data and the high resolution (several eV) soft X-ray 
spectrometer which allows us to study the gas 
kinematics (bulk motion and turbulent velocity) via the width and energy of 
the emission lines. The kinematics of the gas provides unprecedented
information about the bulk motion; the energy of this motion is in turn
responsible for acceleration of particles to very high energies at
shocks, which is in turn manifested via non-thermal processes, best
studied via sensitive hard X-ray measurements.

Precision cosmology uses astronomical observations to determine the 
large-scale structure and content of the Universe. Studies of clusters 
of galaxies have provided independent measurements of the dark energy 
equation of state and strong evidence for the existence of dark matter. 
Using a variety of techniques (including the growth of structure, the 
baryonic fraction in clusters, and the Sunyaev-Zel'dovich effect) 
a well-constructed survey of clusters of galaxies, with the necessary 
supporting data\cite{Ref:Rapetti}, can provide precise measurements of 
cosmological parameters, including the amount and properties of dark energy 
and dark matter. The key step is to connect observables (such as flux and 
temperature) to cluster masses. 
Currently, large area cluster surveys are being carried out using the Sunyaev-Zel'dovich Effect by the Atacama Cosmology Telescope, Planck, and the South Pole Telescope to be joined in the near future by the eROSITA X-ray mission.
To reduce the systematic 
uncertainties on the masses inferred from the coarser data from these surveys, 
a training set of precise cluster masses must be obtained. Measurements of bulk 
motion of cluster of galaxies and amounts of non-thermal energies going 
to cosmic-ray acceleration could reduce the ``{\dots} substantial uncertainties 
in the baryonic physics which prevents their use at a high level of precision at 
the present time" (Dark Energy Task Force\cite{Ref:Albrecht}). Line diagnostics 
with energy resolution of 7~eV greatly reduce the uncertainties in the baryonic 
physics by determining the velocity field, any deviations from thermal 
equilibrium, and an accurate temperature for each cluster. Information 
about the non-thermal particle content of clusters can be determined 
via measurements of their Compton upscattering of the CMB:  this is best 
studied via hard X-ray imaging, providing additional clues about the physical 
state of the cluster gas.  

\section*{Acknowledgments}

The authors are deeply grateful for on-going contributions provided by other members in the ASTRO-H team in Japan, the US and Europe.

\end{document}